# Recover Feasible Solutions for SOCP Relaxation of Optimal Power Flow Problems in Mesh Networks

Zhuang Tian, Wenchuan Wu, *Senior Member, IEEE*

*Abstract*—Convex relaxation methods have been studied and used extensively to obtain an optimal solution to the optimal power flow (OPF) problem. Meanwhile, convex relaxed power flow equations are also prerequisites for efficiently solving a wide range of problems in power systems including mixed-integer nonlinear programming (MINLP) and distributed optimization. When the exactness of convex relaxations is not guaranteed, it is important to recover a feasible solution for the convex relaxation methods. This paper presents an alternative convex optimization (ACP) approach that can efficiently recover a feasible solution from the result of second-order cone programming (SOCP) relaxed OPF in mesh networks. The OPF problem is first formulated as a difference-of-convex (DC) programming problem, then efficiently solved by a penalty convex concave procedure (CCP). CCP iteratively linearizes the concave parts of the power flow constraints and solves a convex approximation of the DCP problem. Numerical tests show that the proposed method can find a global or near-global optimal solution to the AC OPF problem, and outperforms those semidefinite programming (SDP) based algorithms.

NOMENCLATURE

A. *Indices and Sets:*

$\Phi_b$   Sets of all buses.
$\Phi_l$   Sets of all lines.
$K(i)$   Sets of buses connected to bus $i$.

B. *Parameters:*

$G_{ij}$   Conductance of branch $ij$.
$B_{ij}$   Susceptance of branch $ij$.
$G_{sh,i}$   Shunt conductance at bus $i$.
$B_{sh,i}$   Shunt susceptance at bus $i$.
$p_i^d$   Active power demand at bus $i$.
$q_i^d$   Reactive power demand at bus $i$.
$p_i^l, p_i^u$   Active power capacity of generator at bus $i$.
$q_i^l, q_i^u$   Reactive power capacity of generator at to bus $i$.
$\theta^u$   Maximum phase angle difference of each branch.
$S^u$   Maximum apparent power of each branch.
$V^l, V^u$   Voltage magnitude limit of each bus.
$s^l, s^u$   Range of $s_{ij}$, $s^l = -\sin\theta^u$, $s^u = \sin\theta^u$
$c^l, c^u$   Range of $c_{ij}$, $c^l = \cos\theta^u$, $c^u = 1$
$K^l, K^u$   Range of $K_{ij}$, $K^l = (V^l)^2 c^l$, $K^u = (V^u)^2$
$L^l, L^u$   Range of $L_{ij}$, $L^l = (V^u)^2 s^l$, $L^u = (V^u)^2 s^u$
$\tau$   Penalty parameters in convex concave procedure.
$\delta_1, \delta_2$   Stopping criterion in convex concave procedure.

C. *Variables:*

$V_i$   Voltage magnitude of bus $i$.
$\theta_i$   Phase angle of bus $i$.
$\theta_{ij}$   Phase angle difference of branch $ij$.
$p_{ij}$   Active power flow from bus $i$ to bus $j$.
$q_{ij}$   Reactive power flow from bus $i$ to bus $j$.
$p_i^g$   Active power provided by generator at bus $i$.
$q_i^g$   Reactive power provided by generator at bus $i$.
$U_i$   Square of $V_i$.
$K_{ij}$   Denotes $V_i V_j \cos\theta_{ij}$.
$L_{ij}$   Denotes $V_i V_j \cos\theta_{ij}$.
$s_{ij}$   Denotes $\sin\theta_{ij}$.
$c_{ij}$   Denotes $\cos\theta_{ij}$.
$\varepsilon$   Slack variables in convex concave procedure.

I. INTRODUCTION

THE AC optimal power flow (OPF) problem is essential for power systems to determine the operation point that best minimizes generation cost, power losses, voltage fluctuations, and other crucial outcomes. It is a typical nonconvex and NP-hard problem, for which the non-convexity mainly lies in the power flow equations. Traditional methods to solve OPF problems in transmission systems include linear approximations, the Newton-Raphson method and some heuristic algorithms, which either lack feasibility or cannot ensure optimality. With the increasing penetration of renewable generations, the OPF problem for power systems has drawn much attention in recent years. MINLP and decentralized optimization problems also require convex formulation of power flow equations so that the problem can be solved efficiently. The convex relaxation of OPF problems was first proposed in [1], [2], and has become an important research topic in the past five years.

Convex relaxation methods mainly include semidefinite

Manuscript received XXXX, 2017. The authors are with the State Key Laboratory of Power Systems, Department of Electrical Engineering, Tsinghua University, Beijing 100084, China (email: wuwench@tsinghua.edu.cn). This work was supported in part by the National Key Research and Development Plan of China (Grant.2016YFB0900400), in part by the National Science Foundation of China (Grant. 51725703)

programming (SDP) relaxation [1] and second-order cone programming (SOCP) relaxation [2]. These methods can find a lower bound of the original minimization problem, and in certain circumstances, a feasible solution of the original problem can be recovered from the solution of convex relaxation methods. For SDP relaxation, if the rank one condition is satisfied, then the zero-duality gap can be guaranteed, hence a feasible solution is sure to be recovered [3]. For SOCP relaxation, a feasible solution can be recovered when the quadratic and arctangent equalities both hold [4], [5]. Under such circumstances, we say the convex relaxation is exact, and its solution is the global optimum of the original OPF problem. Many researchers have devoted efforts toward finding sufficient conditions for ensuring the exactness of, or strengthening, the convex relaxations.

In [6], a sufficient zero-duality-gap condition for SDP was found in resistive networks with active loads if over-satisfaction of the loads was permitted. In [7] and [8], sufficient conditions for SDP in radial and mesh networks were discussed. In [9], the exactness of SDP for mesh networks was related to the modelling of the capacity of a power line. In [10], the sufficient condition for SOCP in radial networks was proposed. If the objective function of the OPF problem is non-increasing in load, and there are no upper limits on load, then the solution of SOCP is exact for radial networks. In [11], three types of sufficient condition were discussed: power injections, voltage magnitudes, and voltage angles. A mild condition that only limits the power injections was proposed. In general, SOCP relaxation is excellent for solving OPF problems in radial networks, besides the sufficient conditions, its tightness can be checked a posteriori for many problems. In [12], the performance of SOCP in mesh networks was further studied. In [13], a cycle-based formulation of angle constraints was proposed to enhance SOCP relaxation. By exploring the fact that angle differences sum up to zero over each cycle, the angle constraints were transformed into bilinear constraints. However, there has not been a method that guarantees the exactness of SOCP relaxation in mesh networks, because the conic relaxation and the angle relaxation must both be exact to ensure the feasibility of the SOCP solution in mesh networks, but the angle constraints are difficult to deal with due to the trigonometric functions.

While the above literatures discussed the exactness of convex relaxation methods, it is still an important issue that how to recover a feasible solution of the original OPF problem when the exactness of convex relaxation is not guaranteed, especially for SOCP relaxation in mesh networks. The motivation for feasibility recovery is to make convex relaxation more practical in sophisticated problems based on OPF, such as MINLP or distributed optimization problems in power systems:

1) In MINLP problems such as transmission line switching [14] or voltage control considering the adjustment of transformer's OLTC (On-Load Tap-Changer) [15], the power flow equations need to be convexified or linearized so that the problem can be to efficiently solved. To obtain a physically meaningful solution, feasibility recovery should be utilized.

2) In distributed optimization problems, the convergence of distributed algorithms, such as ADMM [16], can only be guaranteed for convex problems [17]. In such problems, the convex relaxed power flow equations are employed. So, the solution of distributed OPF must be recovered to a feasible solution to make the strategies practical.

In [9], a penalized SDP method is proposed, the total amount of reactive power was added to the objective to force the rank to become one. In [20], the matrix rank is approximated by a continuous function and penalized in the objective function, then a majorization-minimization method is applied to solve the penalized SDP problem iteratively. In [21], moment relaxations were proposed for the OPF problem as a generalization of SDP relaxation, and had the potential to find a global optimal solution using polynomial optimization theory. Moment relaxations significantly increase the matrix size of semidefinite constraints, which is much more computational inefficient than SDP. In [22], a Laplacian-based approach was proposed to yield near-globally optimal solutions when SDP had a small optimality gap. In [23] and [24], instead of forcing matrix rank to be one, they employed the quadratically constrained quadratic programming (QCQP) formulation of OPF problems, and applied convex concave procedure to deal with the indefinite coefficient matrix. In [25], when the SOCP relaxation is inexact, the OPF problem in radial networks was first formulated as a difference-of-convex programming problem (DCP), then solved as a sequence of convexified penalization problems.

However, there is no method yet available to recover a feasible and optimal or near-optimal solution for the SOCP relaxation in mesh networks. This paper applies the convex concave procedure (CCP) to the OPF problem in mesh networks and recovers a feasible and local optimal solution for SOCP relaxation. CCP is a powerful heuristic method for finding a local optimum of DCP problems [26], which was first introduced in [27] and [28]. It iteratively linearizes the concave parts of all constraints, thus solving a convex approximation of the DCP problem. In [29], penalty CCP was proposed to negate the need for an initial feasible point in the iteration. Penalty CCP usually benefits from a warm-start point, which makes good use of the solution solved by SOCP. The main contributions of this paper include:

1) An alternative convex optimization (ACP) algorithm is proposed that can efficiently recover a feasible solution from the result of SOCP relaxed OPF problem in mesh networks. The ACP algorithm first formulates OPF problem as a DCP problem, then solves the DCP problem by penalty CCP iteratively.

2) The convergence of ACP is proved. After ACP converges, if the slack variables all turn out to be zero, then the solution is guaranteed to be a KKT point of the original OPF problem. It is shown that ACP successfully converges to a KKT point of the original OPF problem in all the test cases.

3) Numerical tests are conducted on several benchmark systems using ACP and compared with other methods aimed to

recover feasible solutions for SDP relaxation. It is shown that the proposed algorithm can find a global or near-global optimal solution within a few iterations, but the other recovery methods may only find worse results. Its computation speed is comparable to SOCP, which is far beyond SDP-based recover methods. Furthermore, an optimal control of OLTC test case is studied to show the availability of ACP in MINLP problems when other methods fail.

The remainder of this paper is organized as follows. Section II describes the original non-convex model of OPF in mesh networks and a tightened SOCP relaxation, and Section III details the DCP formulation of the OPF problem and the ACP algorithm for feasible solution recovery. Section IV outlines the test results of the algorithm using several IEEE test systems, and Section V concludes the paper.

## II. OPF Problem and Convex Relaxation

### A. Original OPF problem

The OPF problem usually consists of convex functions of generator output, denoted by $C_i(p_i^g)$. This is described as:

(Model 1)  $\quad \min \sum C_i(p_i^g) \quad$ (1)

Subject to

1) Branch power flow constraints

$$p_{ij} = G_{ij}V_i^2 - G_{ij}V_iV_j\cos\theta_{ij} - B_{ij}V_iV_j\sin\theta_{ij}, \forall ij \in \Phi_l \quad (2)$$

$$q_{ij} = -B_{ij}V_i^2 + B_{ij}V_iV_j\cos\theta_{ij} - G_{ij}V_iV_j\sin\theta_{ij}, \forall ij \in \Phi_l \quad (3)$$

$$\theta_{ij} = \theta_i - \theta_j, \forall ij \in \Phi_l \quad (4)$$

2) Active and reactive power balance constraints for buses

$$p_i^g - p_i^d = G_{sh,i}V_i^2 + \sum_{j \in K(i)} p_{ij}, \forall i \in \Phi_b \quad (5)$$

$$q_i^g - q_i^d = -B_{sh,i}V_i^2 + \sum_{j \in K(i)} q_{ij}, \forall i \in \Phi_b \quad (6)$$

3) Generator operation constraints

$$p_i^l \leq p_i^g \leq p_i^u, \forall i \in \Phi_b \quad (7)$$

$$q_i^l \leq q_i^g \leq q_i^u, \forall i \in \Phi_b \quad (8)$$

4) Phase angle difference limits

$$-\theta^u \leq \theta_{ij} \leq \theta^u, \forall ij \in \Phi_l \quad (9)$$

5) Branch thermal limits

$$p_{ij}^2 + q_{ij}^2 \leq (S^u)^2, \forall ij \in \Phi_l \quad (10)$$

6) Bus voltage limits

$$V^l \leq V_i \leq V^u, \forall i \in \Phi_b \quad (11)$$

The original formulation of OPF problem is nonconvex and the non-convexity comes from branch power flow constraints (2) and (3). The challenge of non-convexity in realistic power systems OPF also comes from transformer taps, capacitor, etc., which has been discussed in [15] and [30]. So, in this paper, we mainly focus on the nonconvex power flow constraints.

By defining new variables $K_{ij} = V_iV_j\cos\theta_{ij}$, $L_{ij} = V_iV_j\sin\theta_{ij}$ and $U_i = V_i^2$, constraints (2), (3), (5), and (6) can be transformed into an alternative form:

$$p_i^g - p_i^d = G_{sh,i}U_i + \sum_{j \in K(i)} p_{ij}, \forall i \in \Phi_b \quad (12)$$

$$q_i^g - q_i^d = -B_{sh,i}U_i + \sum_{j \in K(i)} q_{ij}, \forall i \in \Phi_b \quad (13)$$

$$p_{ij} = G_{ij}U_i - G_{ij}K_{ij} - B_{ij}L_{ij}, \forall ij \in \Phi_l \quad (14)$$

$$q_{ij} = -B_{ij}U_i + B_{ij}K_{ij} - G_{ij}L_{ij}, \forall ij \in \Phi_l \quad (15)$$

$$K_{ij}^2 + L_{ij}^2 = U_iU_j, \forall ij \in \Phi_l \quad (16)$$

$$\theta_{ij} = \arctan(L_{ij}/K_{ij}), \forall ij \in \Phi_l \quad (17)$$

For the OPF of a radial network, constraints (4) and (17) are not necessary because the optimal solution $K_{ij}$ and $L_{ij}$ will always recover a set of $\theta_{ij}$ and $\theta_{ij}$ that satisfy these two constraints. However, for the OPF of a meshed network, constraints (4) and (17) are necessary to ensure that $\theta_{ij}$ sums to zero over all cycles [13].

Constraint (17) is equivalent to:

$$\sin\theta_{ij}K_{ij} = \cos\theta_{ij}L_{ij}, \forall ij \in \Phi_l \quad (18)$$

By introducing new variables $s_{ij}$, $c_{ij}$, (18) is equivalent to:

$$s_{ij} = \sin\theta_{ij}, \forall ij \in \Phi_l \quad (19)$$

$$c_{ij} = \cos\theta_{ij}, \forall ij \in \Phi_l \quad (20)$$

$$s_{ij}^2 + c_{ij}^2 = 1, \forall ij \in \Phi_l \quad (21)$$

$$s_{ij}K_{ij} = c_{ij}L_{ij}, \forall ij \in \Phi_l \quad (22)$$

With the above transformation, the OPF problem (Model 1) is equivalent to:

(Model 2)  $\quad \min \sum C_i(p_i^g) \quad$ (23)

Subject to

(4), (7)–(16), and (19)–(22)

### B. Tightened SOCP relaxation

The OPF problem (Model 2) is nonconvex due to constraints (16) and (19)–(22). Constraint (16) can be relaxed to a second-order cone constraint [10]:

$$\left\| \begin{matrix} 2K_{ij} \\ 2L_{ij} \\ U_i - U_j \end{matrix} \right\|_2 \leq U_i + U_j, \forall ij \in \Phi_l \quad (24)$$

Constraints (19) and (20) can be relaxed by convex envelopes for sine and cosine functions [31]:

$$s_{ij} \leq \cos(\frac{\theta^u}{2})(\theta_{ij} - \frac{\theta^u}{2}) + \sin(\frac{\theta^u}{2}), \forall ij \in \Phi_l \quad (25)$$

$$s_{ij} \geq \cos(\frac{\theta^u}{2})(\theta_{ij} + \frac{\theta^u}{2}) - \sin(\frac{\theta^u}{2}), \forall ij \in \Phi_l \quad (26)$$

$$c_{ij} \leq 1 - (1-\cos(\theta^u))\theta_{ij}^2/(\theta^u)^2, \forall ij \in \Phi_l \quad (27)$$

$$c_{ij} \geq \cos(\theta^u), \forall ij \in \Phi_l \quad (28)$$

Constraint (21) can be relaxed to:

$$s_{ij}^2 + c_{ij}^2 \leq 1, \forall ij \in \Phi_l \quad (29)$$

For constraint (22), by introducing new variables $m_{ij}$ and $n_{ij}$, it is equivalent to:

$$m_{ij} = s_{ij}K_{ij}, \forall ij \in \Phi_l \quad (30)$$

$$n_{ij} = c_{ij}L_{ij}, \forall ij \in \Phi_l \quad (31)$$

$$m_{ij} = n_{ij}, \forall ij \in \Phi_l \quad (32)$$

Constraints (30) and (31) can be relaxed by McCormick envelopes for bilinear terms [32]:

$$m_{ij} \geq s^l K_{ij} + s_{ij} K^l - s^l K^l, \forall ij \in \Phi_l \quad (33)$$

$$m_{ij} \geq s^u K_{ij} + s_{ij} K^u - s^u K^u, \forall ij \in \Phi_l \quad (34)$$

$$m_{ij} \leq s^l K_{ij} + s_{ij} K^u - s^l K^u, \forall ij \in \Phi_l \quad (35)$$

$$m_{ij} \leq s^u K_{ij} + s_{ij} K^l - s^u K^l, \forall ij \in \Phi_l \quad (36)$$

$$n_{ij} \geq c^l L_{ij} + c_{ij} L^l - c^l L^l, \forall ij \in \Phi_l \quad (37)$$

$$n_{ij} \geq c^u L_{ij} + c_{ij} L^u - c^u L^u, \forall ij \in \Phi_l \quad (38)$$

$$n_{ij} \leq c^l L_{ij} + c_{ij} L^u - c^l L^u, \forall ij \in \Phi_l \quad (39)$$

$$n_{ij} \leq c^u L_{ij} + c_{ij} L^l - c^u L^l, \forall ij \in \Phi_l \quad (40)$$

Thus, the tightened SOCP relaxed OPF problem (SOCPT) is expressed as follows:

(Model 3) $\quad \min \sum C_i(p_i^g) \quad (41)$

subject to

(4), (7)–(15), (24)–(29), and (32)–(40)

### III. FEASIBLE SOLUTION RECOVERY ALGORITHM

#### A. Difference-of-convex formulation

The relaxation exactness is barely guaranteed by convex relaxed Model 3, because equality (19), (20) and (22) are hard to be satisfied by convex envelopes (25)-(28) and McCormick relaxation (33)–(40), so that a feasible solution can not be recovered from the solution of Model 3 directly. On the other hand, if the bilinear constraints (16), (21) and (22) are satisfied and the trigonometric functions (19) and (20) are well approximated, then the solution will be feasible to the original OPF problem.

In order to satisfy the bilinear constraints (16), (21) and (22), we formulate them as difference-of-convex constraints, which can be solved by DCP algorithms effectively. In such formulation, the equalities are not easy to loosen as in convex relaxation. Take constraint (22) as an example, it can be written in an alternative form:

$$(s_{ij} + K_{ij})^2 - (s_{ij} - K_{ij})^2 = (c_{ij} + L_{ij})^2 - (c_{ij} - L_{ij})^2 \quad (42)$$

which is equivalent to two difference-of-convex constraints:

$$(s_{ij} + K_{ij})^2 + (c_{ij} - L_{ij})^2 - (s_{ij} - K_{ij})^2 - (c_{ij} + L_{ij})^2 \leq 0 \quad (43)$$

$$(s_{ij} - K_{ij})^2 + (c_{ij} + L_{ij})^2 - (s_{ij} + K_{ij})^2 - (c_{ij} - L_{ij})^2 \leq 0 \quad (44)$$

Considering constraints (16), (21), and (22), we can define the following convex functions:

$$f_{ij,1}(x) = (U_i + U_j)^2 \quad (45)$$

$$f_{ij,2}(x) = 1 \quad (46)$$

$$f_{ij,3}(x) = (s_{ij} + K_{ij})^2 + (c_{ij} - L_{ij})^2 \quad (47)$$

$$g_{ij,1}(x) = (2K_{ij})^2 + (2L_{ij})^2 + (U_i - U_j)^2 \quad (48)$$

$$g_{ij,2}(x) = s_{ij}^2 + c_{ij}^2 \quad (49)$$

$$g_{ij,3}(x) = (s_{ij} - K_{ij})^2 + (c_{ij} + L_{ij})^2 \quad (50)$$

Thus, constraints (16), (21), and (22) can be expressed as difference-of-convex constraints:

$$g_{ij,m}(x) - f_{ij,m}(x) \leq 0, \forall ij \in \Phi_l, m = 1,2,3 \quad (51)$$

$$f_{ij,m}(x) - g_{ij,m}(x) \leq 0, \forall ij \in \Phi_l, m = 1,2,3 \quad (52)$$

For the precise approximation of trigonometric functions (19) and (20), sixth-order Taylor expansion of the cosine function is utilized as follows:

$$c_{ij} = 1 - \theta_{ij}^2/2 + \theta_{ij}^4/24 - \theta_{ij}^6/720 \quad (53)$$

By introducing $\alpha_{ij} = \theta_{ij}^2$, $\beta_{ij} = \theta_{ij}^4$, $\gamma_{ij} = \theta_{ij}^6$, and define the following convex functions:

$$f_{ij,4}(x) = \alpha_{ij}, g_{ij,4}(x) = \theta_{ij}^2 \quad (54)$$

$$f_{ij,5}(x) = \beta_{ij}, g_{ij,5}(x) = \alpha_{ij}^2 \quad (55)$$

$$f_{ij,6}(x) = (\alpha_{ij} + \gamma_{ij})^2, g_{ij,6}(x) = (\alpha_{ij} - \gamma_{ij})^2 + (2\beta_{ij})^2 \quad (56)$$

(53) can be written in a difference-of-convex form:

$$c_{ij} = 1 - \alpha_{ij}/2 + \beta_{ij}/24 - \gamma_{ij}/720 \quad (57)$$

$$g_{ij,m}(x) - f_{ij,m}(x) \leq 0, \forall ij \in \Phi_l, m = 4,5,6 \quad (58)$$

$$f_{ij,m}(x) - g_{ij,m}(x) \leq 0, \forall ij \in \Phi_l, m = 4,5,6 \quad (59)$$

Here, only the cosine function (20) needs to be approximated because equation (21) is satisfied by difference-of-convex constraints (51) and (52).

It should be noted that in realistic power systems operation, $\theta^u$ is usually very small, i.e., less than $5°$. In this situation, directly using $\sin(\theta_{ij}) = \theta_{ij}$ will also be a good approximation.

With difference-of-convex formulation (51), (52) for bilinear terms and accurate approximations (57)-(59) for trigonometric functions, OPF problem (Model 2) can be formulated as a DCP problem:

(Model 4) $\quad \min \sum C_i(p_i^g) \quad (60)$

subject to

(4), (7)–(15), (51), (52), (57)-(59)

Here, the constraints in (51) corresponding to $m = 1,2$ and the constraints in (58) are convex. However, the constraints in (52) and (59) are nonconvex.

Comparing Model 2 and Model 4, the only different constraints are (19), (20) and (57)-(59). Whether the solution of Model 4 is feasible to Model 2 is depended on the quality of approximations (57)-(59). The approximation error between (57) and (19), denoted by $|\cos\theta_{ij} - c_{ij}|$, is less than $10^{-10}$ when $\theta_{ij} < 10°$ and less than $10^{-3}$ when $\theta_{ij} < 90°$, which applies to power systems in most cases.

#### B. Penalty convex-concave procedure

The OPF problem is formed as a DCP problem (Model 4) in part A; thus, penalty CCP can be applied to find a local optimum of Model 4. The procedure for penalty CCP is in two parts:

1) Tighten the difference-of-convex constraints via partial linearization. For example, $g_{ij,m}(x)$ can be linearized around point $x^{(0)}$ as

$$\hat{g}_{ij,m}(x, x^{(0)}) = g_{ij,m}(x^{(0)}) - \nabla g_{ij,m}(x^{(0)})^T(x - x^{(0)}) \quad (61)$$

Since $g_{ij,m}(x)$ is convex, we have $g_{ij,m}(x) \geq \hat{g}_{ij,m}(x, x^{(0)})$, and (52) can be tightened into a convex constraint

$$f_{ij,m}(x) - \hat{g}_{ij,m}(x, x^{(k)}) \leq 0 \tag{62}$$

Constraint (62) reduces the feasible region of the original problem, which may lead to infeasibility, so part 2) is needed.

2) Relax constraint (62) by adding slack variables

$$f_{ij,m}(x) - \hat{g}_{ij,m}(x, x^{(k)}) \leq \varepsilon \tag{63}$$

and penalize the sum of constraint violations in the objective function. By doing so, the problem is always feasible.

The steps for ACP are described in Algorithm 1.

**Algorithm 1: ACP**

**Initialization:**
1. Set the value of $x^{(0)}$ to the solution of Model 3.
2. Set $\tau^{(0)} > 0$, $\tau_{\max}$, $\mu > 1$ and $k = 0$.

**Repeat**
1. Convexify
$$\hat{g}_{ij,m}(x, x^{(k)}) = g_{ij,m}(x^{(k)}) - \nabla g_{ij,m}(x^{(k)})^T (x - x^{(k)}), m = 1, \ldots, 6$$
$$\hat{f}_{ij,3}(x, x^{(k)}) = f_{ij,3}(x^{(k)}) - \nabla f_{ij,3}(x^{(k)})^T (x - x^{(k)})$$

2. Set the value of $x^{(k+1)}$ to the solution of

(Model 5) $\quad \min \sum C_i(p_i^g) + \tau^{(k)} \sum_{ij \in \Phi_l} \sum_{m=1}^{7} \varepsilon_{ij,m}^{(k)} \tag{64}$

subject to

(4), (7)–(15), (51), (57)

$$f_{ij,m}(x) - \hat{g}_{ij,m}(x, x^{(k)}) \leq \varepsilon_{ij,m}^{(k)}, \forall ij \in \Phi_l, m = 1, \ldots, 6 \tag{65}$$

$$g_{ij,3}(x) - \hat{f}_{ij,3}(x, x^{(k)}) \leq \varepsilon_{ij,7}^{(k)}, \forall ij \in \Phi_l \tag{66}$$

$$g_{ij,m}(x) - f_{ij,m}(x) \leq 0, \forall ij \in \Phi_l, m = 1, 2, 4, 5, 6 \tag{67}$$

$$\varepsilon_{ij,m}^{(k)} \geq 0, \forall ij \in \Phi_l, m = 1, \ldots, 7 \tag{68}$$

3. Update $\tau^{(k+1)} = \min(\mu \tau^{(k)}, \tau_{\max})$
4. Update iteration $k = k + 1$.

**Until the stopping criterion is satisfied.**

Model 5 is convex, and the number of second-order cone constraints, as well as quadratic constraints, grows linearly with the number of branches in the system. So, Model 5 can be solved easily and quickly using software packages such as Gourbi, CPLEX, or MOSEK. As for the convergence of ACP, it can be proved that the objective value will converge.

*Proposition 1:* The objective value of Model 5 will converge.

*Proof:* Suppose $(x^{(k)}, \varepsilon^{(k)})$ is the optimal solution to Model 5 in iteration $k$.

We will first prove that $(x^{(k)}, \varepsilon^{(k)})$ is a feasible solution to Model 5 in iteration $k+1$. Since the different constraints in iteration $k$ and $k+1$ are (65) and (66), it suffices to show that $(x^{(k)}, \varepsilon^{(k)})$ satisfies (65) and (66) in iteration $k+1$. That is to prove:

$$f_{ij,m}(x^{(k)}) - \hat{g}_{ij,m}(x^{(k)}, x^{(k)}) \leq \varepsilon_{ij,m}^{(k)}, m = 1, 2, 3 \tag{69}$$

$$g_{ij,3}(x^{(k)}) - \hat{f}_{ij,3}(x^{(k)}, x^{(k)}) \leq \varepsilon_{ij,m}^{(k)} \tag{70}$$

As $(x^{(k)}, \varepsilon^{(k)})$ is the optimal solution to iteration $k$, we have

$$f_{ij,m}(x^{(k)}) - \hat{g}_{ij,m}(x^{(k)}, x^{(k-1)}) \leq \varepsilon_{ij,m}^{(k)} \tag{71}$$

The convexity of $g_{ij,m}(x)$ gives

$$f_{ij,m}(x^{(k)}) - g_{ij,m}(x^{(k)}) \leq f_{ij,m}(x^{(k)}) - \hat{g}_{ij,m}(x^{(k)}, x^{(k-1)}) \tag{72}$$

Substituting $g_{ij,m}(x^{(k)}) = \hat{g}_{ij,m}(x^{(k)}, x^{(k)})$ into (72), together with (71), we have

$$f_{ij,m}(x^{(k)}) - \hat{g}_{ij,m}(x^{(k)}, x^{(k)}) \leq \varepsilon_{ij,m}^{(k)} \tag{73}$$

Thus, (69) holds, and (70) can be proved in a similar way. So $(x^{(k)}, \varepsilon^{(k)})$ is a feasible solution to Model 5 in iteration $k+1$.

We will now show that the objective value is non-increasing. Let $v^{(k)}(x, \varepsilon)$ denote the objective function of Model 5 in iteration $k$. When $k > \log_\mu(\tau_{\max} / \tau^{(0)})$, $\tau^{(k)} = \tau_{\max}$, the objective function (64) will not change, which means

$$v^{(k+1)}(x, \varepsilon) = v^{(k)}(x, \varepsilon) \tag{74}$$

Since $(x^{(k)}, \varepsilon^{(k)})$ is a feasible solution to Model 5 in iteration $k+1$ and $(x^{(k+1)}, \varepsilon^{(k+1)})$ is the optimal solution, it follows that

$$v^{(k+1)}(x^{(k+1)}, \varepsilon^{(k+1)}) \leq v^{(k+1)}(x^{(k)}, \varepsilon^{(k)}) = v^{(k)}(x^{(k)}, \varepsilon^{(k)}) \tag{75}$$

This shows that the objective value is non-increasing. Since both $\sum C_i(p_i^g)$ and $\varepsilon$ have lower bounds, the objective value will converge, which completes the proof.

According to *Proposition 1*, the stopping criterion of ACP can be chosen as:

$$\frac{v^{(k)}(x^{(k)}, \varepsilon^{(k)}) - v^{(k+1)}(x^{(k+1)}, \varepsilon^{(k+1)})}{v^{(k+1)}(x^{(k)}, \varepsilon^{(k)})} \leq \delta_1 \tag{76}$$

when $k > \log_\mu(\tau_{\max} / \tau^{(0)})$, which indicates that the objective value converges. Or $\sum_{ij \in \Phi_l} \sum_{m=1}^{7} \varepsilon_{ij,m}^{(k)} \leq \delta_2 \approx 0$, which means $x^{(k)}$ is already feasible for Model 4.

When ACP converges, if the slack variables all turn out to be zero, then the solution of Model 5 is a feasible solution to Model 4. In this situation, the feasible set of Model 5 is a subset of Model 4, and the solution will be a local optimum to Model 4, which also means it is a KKT point to Model 4. As long as the approximations (57)-(59) are accurate enough, the solution will be a KKT point to the original OPF problem.

Although the objective value of Model 5 will converge, it may converge to an infeasible point of the original OPF problem if the slack variables are not equal to zero. The convergence behavior of ACP depends mainly on two points:

1) The penalty parameter $\tau_{\max}$.

$\tau_{\max}$ should not be too small, because this leads easily to nonzero slack variables, nor too large, which may cause numerical problems.

2) The initial point.

A good starting point helps ACP finding a solution that all slack variables equal to zero. Since ACP aims to recover a feasible solution for SOCP relaxation, the initial point is chosen to be the result of convex relaxed OPF Model 3, which is actually a good choice considering both optimality and computation speed. It should be clarified that the initial point for ACP means $x^{(k)}$ in (65) and (66) used for linearization, while initial values for the whole problem is not needed because SOCP and ACP are both convex optimization problems.

It is shown in the test results that, by choosing $\tau_{max}$ and the initial point appropriately, ACP always converges to a feasible point where all the slack variables are equal to zero within a few iterations.

## IV. NUMERICAL RESULTS

In this section, IEEE benchmark test systems were used to demonstrate the effectiveness of the proposed algorithm. First, the nonlinear solver IPOPT was applied to find a local optimum of the original OPF problem. Then, Three SDP-based heuristic models aiming to recover feasible solutions for SDP relaxation were created to show whether they can achieve a feasible solution to the original OPF problem. Finally, the proposed ACP algorithm was tested to show its ability to recover a feasible solution for SOCP relaxation of the original OPF problem and compared with the other heuristic methods.

The SDP-based heuristic methods used in this paper are described as follows:
1) Penalized SDP Relaxation in [9] (PSDP1). In this method, the total amount of reactive power is added to the objective function to force the matrix rank to become one.
2) Penalized SDP Relaxation in [20] (PSDP2). In this method, the matrix rank is approximated by a continuous function and penalized in the objective function. The penalized SDP problem is solved by majorization-minimization method iteratively.
3) Difference-of-convex Programming in [25] (DSDP). In this method, the matrix rank one constraints are formulated as difference-of-convex inequalities, thus solved by convex -concave procedure iteratively. When the rank one equality is satisfied, SDP is identical to QCQP, so that DSDP can be regarded as a specific formulation of the method in [23].

The ACP algorithm, along with PSDP1, PSDP2, and DSDP, was implemented using YALMIP and MATLAB R2016a software. The SDP relaxation was implemented using sparse technique [34]. All the models were solved by MOSEK. Numerical tests were performed on a computer with an Intel® Core™ i5 (2.30 GHz) processor and 8 GB RAM. The original OPF problem was solved by MATPOWER using an IPOPT solver.

### A. 9-bus test system

The IEEE 9-bus system consists of three generators and nine branches. The branch, bus, generator and generator cost data of the system were taken from MATPOWER. There were three generators connected to buses 1, 2, and 3, and the total real and reactive power capacity were 0 to 820 MW and -900 to 900 MVar, respectively. The voltage of bus 1 was set to $1.0\angle 0°$. The lower and upper bounds of system bus voltages were 0.9 p.u. and 1.1 p.u., and the maximum phase angle difference was 10°.

The solution of MATPOWER is assumed to be a benchmark solution to the original OPF problem, and the sub-optimality gap of the heuristic methods are defined as:

$$\text{Gap} = \frac{v^{other} - v^{MP}}{|v^{MP}|} \times 100\% \qquad (77)$$

where $v^{MP}$ is the objective value of the MATPOWER solution and $v^{other}$ is the objective value of ACP, PSDP1, PSDP2, and DSDP.

To demonstrate the effectiveness of the proposed algorithm, different scenarios were tested:
1) Generation cost minimization (congested operation)

In this test case, we considered the cost of generators under congested operating conditions. The maximum apparent power for each branch was set to 120MVA, and MATPOWER indicated that three branches were reaching its limit. Table I shows that ACP could recover a feasible solution to the original OPF problem from the result of tightened SOCP relaxation, as well as PSDP1, PSDP2, and DSDP, which were able to recover feasible solutions from SDP relaxations. While ACP and PSDP1 reached zero sub-optimality gap in this case, PSDP2 only recovered a near-global solution and DSDP recovered solutions far from global optimum.

The computation time for each method is listed in Table I. For the iterative methods, the first iteration was SOCPT or SDP relaxation that aimed to obtain an initial point for the heuristic methods. It can be observed that among these methods, ACP consumed much less computation time than the other three methods, because in each iteration, ACP solved a SOCP optimization problem, while PSDP2 and DSDP solved a SDP optimization problem. Although PSDP1 only needed to solve SDP once, the computation burden of SDP was much larger than SOCP even in a single iteration.

Fig. 1 shows the objective values and sums of slack variables generated by ACP in each iteration: the parameters were set to $\tau^{(0)} = \tau_{max} = 10^5$ and $\delta_1 = 10^{-6}$. It can also be seen that ACP converged in three iterations with the sum of slacks converged to zero, and the objective value generated by the ACP is nonincreasing. Iteration 0 performed a SOCPT relaxation to obtain an initial point for ACP, which was not part of the ACP nonincreasing sequence.

TABLE I
NUMERICAL RESULTS OF 9-BUS SYSTEM

| | Obj. Value | Gap (%) | Rank | Iteration | Solver time (s) |
|---|---|---|---|---|---|
| MP | 5412.98 | - | - | - | 0.75 |
| ACP | 5412.98 | 0.00 | - | 4 | 0.18 |
| PSDP1 | 5413.38 | 0.01 | 1 | - | 1.09 |
| PSDP2 | 5412.98 | 0.00 | 1 | 2 | 3.25 |
| DSDP | 5430.72 | 0.33 | 1 | 5 | 5.38 |

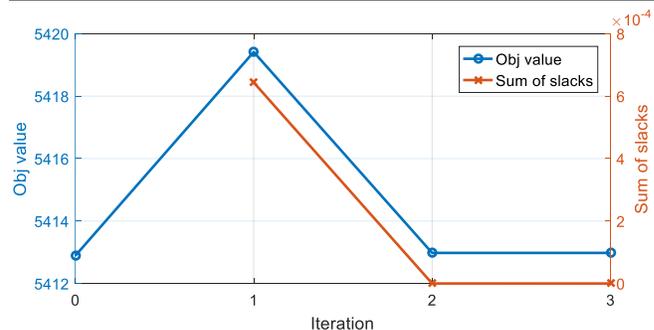

Fig. 1. Convergence behavior of ACP for generation cost minimization

2) Generation cost minimization (large phase angle difference)

The feasibility of ACP is depended on the accuracy of approximations (57)-(59), to test the behavior of ACP when there exists large phase angle differences in power systems, the reactance of branch 1-4 is changed form 0.0576p.u. to 0.576p.u., so that the phase angle difference across this branch can be as large as 33.25°. In this test case, ACP converged in three iterations with slack variables all converged to zero and obtained the same objective as MATPOWER. The voltage magnitude and phase angle data of each bus are shown in Table II. Comparing the results of MATPOWER and ACP, the mismatches in voltage magnitudes and phase angles were both very small, which proved that the solution recovered by ACP was a feasible solution.

TABLE II
OPTIMAL VOLTAGE OF 9-BUS SYSTEM

| Bus | MATPOWER | | ACP | |
|---|---|---|---|---|
| | V (p.u.) | θ (degree) | V (p.u.) | θ (degree) |
| 1 | 1.000 | 0 | 1.000 | 0 |
| 2 | 1.100 | -26.310 | 1.100 | -26.305 |
| 3 | 1.100 | -28.267 | 1.100 | -28.262 |
| 4 | 0.931 | -33.255 | 0.931 | -33.251 |
| 5 | 0.942 | -35.485 | 0.942 | -35.482 |
| 6 | 1.054 | -31.047 | 1.054 | -31.043 |
| 7 | 1.030 | -32.903 | 1.030 | -32.899 |
| 8 | 1.047 | -30.568 | 1.047 | -30.564 |
| 9 | 0.926 | -36.330 | 0.926 | -36.327 |
| Maximum relative error (%) | | | 0.00 | 0.02 |

3) Loss minimization (with transformer)

In this test case, the OLTC of transformer was considered to demonstrate the effectiveness of ACP in MINLP. A transformer was added to line 1-4, the turns ratio was between 0.9 to 1.1 with 0.2 per tap. Exact linearization of transformer [16] introduced binary variables to all the OPF models. The power loss and transformer turns ratio obtained by ACP and SOCPT is shown in Table III. The actual power loss is obtained by running power flow with generator output and turns ratio derived from these two methods. Since SOCPT is not exact, its control effect differs from the optimization results. To verify the results of ACP, the turns ratio of line 1-4 is set to 1.1 manually, and MATPOWER obtained the same result as ACP. The convergence behavior of ACP is shown in Fig. 2, with $\tau^{(0)} = \tau_{max} = 10$ and $\delta_2 = 10^{-5}$, ACP converged in three iterations and the slack variables all converged to zero.

In the MINLP test case, only ACP could recover a feasible solution, whereas the original mixed-integer nonconvex OPF model could not be solved by IPOPT, so well as mixed-integer SDP problems could not be solved by Mosek.

TABLE III
NUMERICAL RESULTS OF 9-BUS SYSTEM

| | Obj. Value | $t_{4-1}$ | Actual Power Loss |
|---|---|---|---|
| ACP | 2.885 | 1.1 | 2.885 |
| SOCPT | 2.875 | 1.1 | 2.886 |

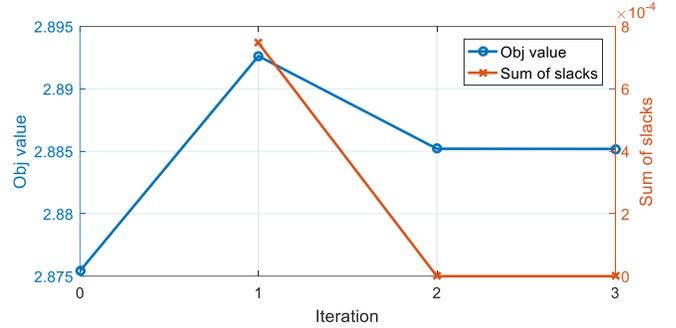

Fig. 2. Convergence behavior of ACP for loss minimization with OLTC

### B. General test systems

The sub-optimality gap and computation time are listed in Table IV. Five test systems and two types of objective function were considered. Among the tested methods, ACP, PSDP1, PSDP2 and DSDP could find feasible solutions as good as MATPOWER in nine, seven, one, four of the ten test cases, respectively. Although none of the heuristic methods are guaranteed to recovery a feasible solution, the test results showed that ACP converged to feasible solutions with slack variables all equal to zero in all the test cases, which were also KKT points of the original OPF problem. On the other hand, the solution recovered by PSDP1 might not be a KKT point, and its behavior was highly sensitive to the penalty parameter, which led to worse results than ACP.

It needs to be pointed out that although we have shown the results and computation time of MATPOWER as reference, ACP is not intended to outperform any state-of-the-art OPF solvers. Instead, ACP seeks to recover a feasible for SOCP relaxation through convex optimization. With this method, MINLP or distributed optimization problems incorporating convex relaxed power flow equations can obtain a feasible and global or near-global optimal solution, so that the control effects will be guaranteed.

TABLE IV
NUMERICAL RESULTS OF IEEE BENCHMARK SYSTEMS

| Test case | Ref. Obj. Value (MATPOWER) | Sub-optimality gap (%) | | | | Solver Time (s) | | | | |
|---|---|---|---|---|---|---|---|---|---|---|
| | | ACP | PSDP1 | PSDP2 | DSDP | MP | ACP | PSDP1 | PSDP2 | DSDP |
| Loss minimization problem (without transformer) (MW) | | | | | | | | | | |
| 9 | 3.546 | 0.00 | 0.00 | 0.11 | 7.87 | 0.69 | 0.11 | 0.80 | 5.61 | 4.65 |
| 14 | 0.635 | 0.00 | 0.00 | 0.00 | 0.00 | 0.70 | 0.15 | 0.83 | 13.00 | 4.08 |
| 30 | 1.777 | 0.00 | 0.00 | 4.46 | 0.00 | 0.93 | 0.18 | 1.35 | 30.42 | 6.79 |
| 57 | 12.148 | 0.00 | 0.00 | 1.20 | 0.12 | 0.80 | 0.41 | 2.52 | 1883.82 | 26.05 |
| 118 | 10.667 | 0.00 | 0.00 | * | 0.83 | 1.09 | 0.92 | 10.96 | >4×10⁵ [20] | 55.13 |
| Generation cost minimization problem (congested operation) ($/h) | | | | | | | | | | |
| 9 | 5329.53 | 0.00 | 0.01 | 0.00 | 0.33 | 0.75 | 0.18 | 1.09 | 3.25 | 5.38 |
| 14 | 9252.28 | 0.00 | 0.00 | × | 0.00 | 0.77 | 0.32 | 0.94 | 5.48 | 4.28 |
| 30 | 582.79 | 0.00 | 0.07 | 0.02 | 0.14 | 1.13 | 5.31 | 2.23 | 58.05 | 38.46 |
| 57 | 43697.64 | 0.00 | 0.00 | × | 0.00 | 0.76 | 1.46 | 4.08 | 1945.05 | 14.5 |
| 118 | 134007.40 | 0.01 | × | * | 0.12 | 1.06 | 2.73 | 10.26 | >4×10⁵ [20] | 62.99 |

×– Infeasible solution, *– numerical problems.

## V. Conclusion

In this paper, an ACP algorithm was proposed to recover a global or near-global optimal solution for SOCP relaxed OPF problem in mesh networks when the convex relaxation method is not exact. The OPF problem was first formulated as a DCP problem to maintain equality in the nonconvex power flow equations, then solved efficiently by penalty CCP. A tightened SOCP relaxation of the OPF problem in mesh networks was also proposed to provide a good initial point for the ACP algorithm. Numerical results showed that the proposed algorithm could recover global or near-global optimal solutions for SOCP relaxation with various objective functions and generally performed better than the SDP-based recovery methods in solution quality. The computational efficiency of the proposed algorithm was comparable to SOCP, which was far beyond the SDP-based methods.

Since every iteration of ACP is a convex optimization problem, the proposed method is suitable for more complicated optimization problems in power systems such as MINLP or distributed control optimization which require a convex formulation of power flow equations. We have demonstrated the capability of ACP in MINLP with a test case considering transformer turns ratio in this paper. The application of ACP in other problems deserve further investigation in our future research.